
%
\catcode`@=11 
%
%
%

\font\fourteenrm=cmr10 scaled\magstep2
\font\twelverm=cmr10 scaled\magstep1
\font\ninerm=cmr9            \font\sixrm=cmr6

\font\fourteenbf=cmbx10 scaled\magstep2
\font\twelvebf=cmbx10 scaled\magstep1
\font\ninebf=cmbx9            \font\sixbf=cmbx6
\font\seventeeni=cmmi10 scaled\magstep3     \skewchar\seventeeni='177
\font\fourteeni=cmmi10 scaled\magstep2      \skewchar\fourteeni='177
\font\twelvei=cmmi10 scaled\magstep1        \skewchar\twelvei='177
\font\ninei=cmmi9                           \skewchar\ninei='177
\font\sixi=cmmi6                            \skewchar\sixi='177
\font\seventeensy=cmsy10 scaled\magstep3    \skewchar\seventeensy='60
\font\fourteensy=cmsy10 scaled\magstep2     \skewchar\fourteensy='60
\font\twelvesy=cmsy10 scaled\magstep1       \skewchar\twelvesy='60
\font\ninesy=cmsy9                          \skewchar\ninesy='60
\font\sixsy=cmsy6                           \skewchar\sixsy='60

\font\fourteenex=cmex10 scaled\magstep2
\font\twelveex=cmex10 scaled\magstep1

\font\fourteensl=cmsl10 scaled\magstep2
\font\twelvesl=cmsl10 scaled\magstep1
\font\ninesl=cmsl9

\font\fourteenit=cmti10 scaled\magstep2
\font\twelveit=cmti10 scaled\magstep1
\font\twelvett=cmtt10 scaled\magstep1
\font\twelvecp=cmcsc10 scaled\magstep1
\font\tencp=cmcsc10
\newfam\cpfam
%
%
\newcount\f@ntkey            \f@ntkey=0
\def\samef@nt{\relax \ifcase\f@ntkey \rm \or\oldstyle \or\or
         \or\it \or\sl \or\bf \or\tt \or\caps \fi }
\def\fourteenpoint{\relax
    \textfont0=\fourteenrm          \scriptfont0=\tenrm
    \scriptscriptfont0=\sevenrm
     \def\rm{\fam0 \fourteenrm \f@ntkey=0 }\relax
    \textfont1=\fourteeni           \scriptfont1=\teni
    \scriptscriptfont1=\seveni
     \def\oldstyle{\fam1 \fourteeni\f@ntkey=1 }\relax
    \textfont2=\fourteensy          \scriptfont2=\tensy
    \scriptscriptfont2=\sevensy
    \textfont3=\fourteenex     \scriptfont3=\fourteenex
    \scriptscriptfont3=\fourteenex
    \def\it{\fam\itfam \fourteenit\f@ntkey=4 }\textfont\itfam=\fourteenit
    \def\sl{\fam\slfam \fourteensl\f@ntkey=5 }\textfont\slfam=\fourteensl
    \scriptfont\slfam=\tensl
    \def\bf{\fam\bffam \fourteenbf\f@ntkey=6 }\textfont\bffam=\fourteenbf
    \scriptfont\bffam=\tenbf     \scriptscriptfont\bffam=\sevenbf
    \def\tt{\fam\ttfam \twelvett \f@ntkey=7 }\textfont\ttfam=\twelvett
    \h@big=11.9\p@{} \h@Big=16.1\p@{} \h@bigg=20.3\p@{} \h@Bigg=24.5\p@{}
    \def\caps{\fam\cpfam \twelvecp \f@ntkey=8 }\textfont\cpfam=\twelvecp
    \setbox\strutbox=\hbox{\vrule height 12pt depth 5pt width\z@}
    \samef@nt}
\def\twelvepoint{\relax
    \textfont0=\twelverm          \scriptfont0=\ninerm
    \scriptscriptfont0=\sixrm
     \def\rm{\fam0 \twelverm \f@ntkey=0 }\relax
    \textfont1=\twelvei           \scriptfont1=\ninei
    \scriptscriptfont1=\sixi
     \def\oldstyle{\fam1 \twelvei\f@ntkey=1 }\relax
    \textfont2=\twelvesy          \scriptfont2=\ninesy
    \scriptscriptfont2=\sixsy
    \textfont3=\twelveex          \scriptfont3=\twelveex
    \scriptscriptfont3=\twelveex
    \def\it{\fam\itfam \twelveit \f@ntkey=4 }\textfont\itfam=\twelveit
    \def\sl{\fam\slfam \twelvesl \f@ntkey=5 }\textfont\slfam=\twelvesl
    \scriptfont\slfam=\ninesl
    \def\bf{\fam\bffam \twelvebf \f@ntkey=6 }\textfont\bffam=\twelvebf
    \scriptfont\bffam=\ninebf     \scriptscriptfont\bffam=\sixbf
    \def\tt{\fam\ttfam \twelvett \f@ntkey=7 }\textfont\ttfam=\twelvett
    \h@big=10.2\p@{}
    \h@Big=13.8\p@{}
    \h@bigg=17.4\p@{}
    \h@Bigg=21.0\p@{}
    \def\caps{\fam\cpfam \twelvecp \f@ntkey=8 }\textfont\cpfam=\twelvecp
    \setbox\strutbox=\hbox{\vrule height 10pt depth 4pt width\z@}
    \samef@nt}
\def\tenpoint{\relax
    \textfont0=\tenrm          \scriptfont0=\sevenrm
    \scriptscriptfont0=\fiverm
    \def\rm{\fam0 \tenrm \f@ntkey=0 }\relax
    \textfont1=\teni           \scriptfont1=\seveni
    \scriptscriptfont1=\fivei
    \def\oldstyle{\fam1 \teni \f@ntkey=1 }\relax
    \textfont2=\tensy          \scriptfont2=\sevensy
    \scriptscriptfont2=\fivesy
    \textfont3=\tenex          \scriptfont3=\tenex
    \scriptscriptfont3=\tenex
    \def\it{\fam\itfam \tenit \f@ntkey=4 }\textfont\itfam=\tenit
    \def\sl{\fam\slfam \tensl \f@ntkey=5 }\textfont\slfam=\tensl
    \def\bf{\fam\bffam \tenbf \f@ntkey=6 }\textfont\bffam=\tenbf
    \scriptfont\bffam=\sevenbf     \scriptscriptfont\bffam=\fivebf
    \def\tt{\fam\ttfam \tentt \f@ntkey=7 }\textfont\ttfam=\tentt
    \def\caps{\fam\cpfam \tencp \f@ntkey=8 }\textfont\cpfam=\tencp
    \setbox\strutbox=\hbox{\vrule height 8.5pt depth 3.5pt width\z@}
    \samef@nt}
%
%
%
%
\newdimen\h@big  \h@big=8.5\p@
\newdimen\h@Big  \h@Big=11.5\p@
\newdimen\h@bigg  \h@bigg=14.5\p@
\newdimen\h@Bigg  \h@Bigg=17.5\p@
\def\big#1{{\hbox{$\left#1\vbox to\h@big{}\right.\n@space$}}}
\def\Big#1{{\hbox{$\left#1\vbox to\h@Big{}\right.\n@space$}}}
\def\bigg#1{{\hbox{$\left#1\vbox to\h@bigg{}\right.\n@space$}}}
\def\Bigg#1{{\hbox{$\left#1\vbox to\h@Bigg{}\right.\n@space$}}}
%
%
%
\normalbaselineskip = 20pt plus 0.2pt minus 0.1pt
\normallineskip = 1.5pt plus 0.1pt minus 0.1pt
\normallineskiplimit = 1.5pt
\newskip\normaldisplayskip
\normaldisplayskip = 20pt plus 5pt minus 10pt
\newskip\normaldispshortskip
\normaldispshortskip = 6pt plus 5pt
\newskip\normalparskip
\normalparskip = 6pt plus 2pt minus 1pt
\newskip\skipregister
\skipregister = 5pt plus 2pt minus 1.5pt
\newif\ifsingl@    \newif\ifdoubl@
\newif\iftwelv@    \twelv@true
\def\singlespace{\singl@true\doubl@false\spaces@t}
\def\doublespace{\singl@false\doubl@true\spaces@t}
\def\normalspace{\singl@false\doubl@false\spaces@t}
\def\Tenpoint{\tenpoint\twelv@false\spaces@t}
\def\Twelvepoint{\twelvepoint\twelv@true\spaces@t}
\def\spaces@t{\relax%
 \iftwelv@ \ifsingl@\subspaces@t3:4;\else\subspaces@t29:31;\fi%
 \else \ifsingl@\subspaces@t3:5;\else\subspaces@t4:5;\fi \fi%
 \ifdoubl@ \multiply\baselineskip by 5%
 \divide\baselineskip by 4 \fi \unskip}
\def\subspaces@t#1:#2;{
      \baselineskip = \normalbaselineskip
      \multiply\baselineskip by #1 \divide\baselineskip by #2
      \lineskip = \normallineskip
      \multiply\lineskip by #1 \divide\lineskip by #2
      \lineskiplimit = \normallineskiplimit
      \multiply\lineskiplimit by #1 \divide\lineskiplimit by #2
      \parskip = \normalparskip
      \multiply\parskip by #1 \divide\parskip by #2
      \abovedisplayskip = \normaldisplayskip
      \multiply\abovedisplayskip by #1 \divide\abovedisplayskip by #2
      \belowdisplayskip = \abovedisplayskip
      \abovedisplayshortskip = \normaldispshortskip
      \multiply\abovedisplayshortskip by #1
        \divide\abovedisplayshortskip by #2
      \belowdisplayshortskip = \abovedisplayshortskip
      \advance\belowdisplayshortskip by \belowdisplayskip
      \divide\belowdisplayshortskip by 2
      \smallskipamount = \skipregister
      \multiply\smallskipamount by #1 \divide\smallskipamount by #2
      \medskipamount = \smallskipamount \multiply\medskipamount by 2
      \bigskipamount = \smallskipamount \multiply\bigskipamount by 4 }
\def\normalbaselines{ \baselineskip=\normalbaselineskip
   \lineskip=\normallineskip \lineskiplimit=\normallineskip
   \iftwelv@\else \multiply\baselineskip by 4 \divide\baselineskip by 5
     \multiply\lineskiplimit by 4 \divide\lineskiplimit by 5
     \multiply\lineskip by 4 \divide\lineskip by 5 \fi }
\Twelvepoint  
\interlinepenalty=50
\interfootnotelinepenalty=5000
\predisplaypenalty=9000
\postdisplaypenalty=500
\hfuzz=1pt
\vfuzz=0.2pt
%
%
%
\def\pagecontents{
   \ifvoid\topins\else\unvbox\topins\vskip\skip\topins\fi
   \dimen@ = \dp255 \unvbox255
   \ifvoid\footins\else\vskip\skip\footins\footrule\unvbox\footins\fi
   \ifr@ggedbottom \kern-\dimen@ \vfil \fi }
\def\makeheadline{\vbox to 0pt{ \skip@=\topskip
      \advance\skip@ by -12pt \advance\skip@ by -2\normalbaselineskip
      \vskip\skip@ \line{\vbox to 12pt{}\the\headline} \vss
      }\nointerlineskip}
\def\makefootline{\baselineskip = 1.5\normalbaselineskip
                 \line{\the\footline}}
\newif\iffrontpage
\newif\ifletterstyle
\newif\ifp@genum
\def\nopagenumbers{\p@genumfalse}
\def\pagenumbers{\p@genumtrue}
\pagenumbers
\newtoks\paperheadline
\newtoks\letterheadline
\newtoks\letterfrontheadline
\newtoks\lettermainheadline
\newtoks\paperfootline
\newtoks\letterfootline
\newtoks\date
\footline={\ifletterstyle\the\letterfootline\else\the\paperfootline\fi}
\paperfootline={\hss\iffrontpage\else\ifp@genum\tenrm\folio\hss\fi\fi}
\letterfootline={\hfil}
\headline={\ifletterstyle\the\letterheadline\else\the\paperheadline\fi}
\paperheadline={\hfil}
\letterheadline{\iffrontpage\the\letterfrontheadline
     \else\the\lettermainheadline\fi}
\lettermainheadline={\rm\ifp@genum page \ \folio\fi\hfil\the\date}
\def\monthname{\relax\ifcase\month 0/\or January\or February\or
   March\or April\or May\or June\or July\or August\or September\or
   October\or November\or December\else\number\month/\fi}
\date={\monthname\ \number\day, \number\year}
\countdef\pagenumber=1  \pagenumber=1
\def\advancepageno{\global\advance\pageno by 1
   \ifnum\pagenumber<0 \global\advance\pagenumber by -1
    \else\global\advance\pagenumber by 1 \fi \global\frontpagefalse }
\def\folio{\ifnum\pagenumber<0 \romannumeral-\pagenumber
           \else \number\pagenumber \fi }
\def\footrule{\dimen@=\prevdepth\nointerlineskip
   \vbox to 0pt{\vskip -0.25\baselineskip \hrule width 0.35\hsize \vss}
   \prevdepth=\dimen@ }
\newtoks\foottokens
\foottokens={\Tenpoint\singlespace}
\newdimen\footindent
\footindent=24pt
\def\vfootnote#1{\insert\footins\bgroup  \the\foottokens
   \interlinepenalty=\interfootnotelinepenalty \floatingpenalty=20000
   \splittopskip=\ht\strutbox \boxmaxdepth=\dp\strutbox
   \leftskip=\footindent \rightskip=\z@skip
   \parindent=0.5\footindent \parfillskip=0pt plus 1fil
   \spaceskip=\z@skip \xspaceskip=\z@skip
   \Textindent{$ #1 $}\footstrut\futurelet\next\fo@t}
\def\Textindent#1{\noindent\llap{#1\enspace}\ignorespaces}
\def\footnote#1{\attach{#1}\vfootnote{#1}}

\let\footsymbol=\star
\newcount\lastf@@t           \lastf@@t=-1
\newcount\footsymbolcount    \footsymbolcount=0
\newif\ifPhysRev
\def\footsymbolgen{\relax \ifPhysRev \iffrontpage \NPsymbolgen\else
      \PRsymbolgen\fi \else \NPsymbolgen\fi
   \global\lastf@@t=\pageno \footsymbol }
\def\NPsymbolgen{\ifnum\footsymbolcount<0 \global\footsymbolcount=0\fi
   {\iffrontpage \else \advance\lastf@@t by 1 \fi
    \ifnum\lastf@@t<\pageno \global\footsymbolcount=0
     \else \global\advance\footsymbolcount by 1 \fi }
   \ifcase\footsymbolcount \fd@f\star\or \fd@f\dagger\or \fd@f\ast\or
    \fd@f\ddagger\or \fd@f\natural\or \fd@f\diamond\or \fd@f\bullet\or
    \fd@f\nabla\else \fd@f\dagger\global\footsymbolcount=0 \fi }
\def\fd@f#1{\xdef\footsymbol{#1}}
\def\PRsymbolgen{\ifnum\footsymbolcount>0 \global\footsymbolcount=0\fi
      \global\advance\footsymbolcount by -1
      \xdef\footsymbol{\sharp\number-\footsymbolcount} }
\def\space@ver#1{\let\@sf=\empty \ifmmode #1\else \ifhmode
   \edef\@sf{\spacefactor=\the\spacefactor}\unskip${}#1$\relax\fi\fi}
\def\attach#1{\space@ver{\strut^{\mkern 2mu #1} }\@sf\ }
\def\atttach#1{\space@ver{\strut{\mkern 2mu #1} }\@sf\ }
%
%
%
\newcount\chapternumber      \chapternumber=0
\newcount\sectionnumber      \sectionnumber=0
\newcount\equanumber         \equanumber=0
\let\chapterlabel=0
\newtoks\chapterstyle        \chapterstyle={\Number}
\newskip\chapterskip         \chapterskip=\bigskipamount
\newskip\sectionskip         \sectionskip=\medskipamount
\newskip\headskip            \headskip=8pt plus 3pt minus 3pt
\newdimen\chapterminspace    \chapterminspace=15pc
\newdimen\sectionminspace    \sectionminspace=10pc
\newdimen\referenceminspace  \referenceminspace=25pc
\def\chapterreset{\global\advance\chapternumber by 1
   \ifnum\the\equanumber<0 \else\global\equanumber=0\fi
   \sectionnumber=0 \makel@bel}
\def\makel@bel{\xdef\chapterlabel{%
\the\chapterstyle{\the\chapternumber}.}}
\def\sectionlabel{\number\sectionnumber \quad }
\def\alphabetic#1{\count255='140 \advance\count255 by #1\char\count255}
\def\Alphabetic#1{\count255='100 \advance\count255 by #1\char\count255}
\def\Roman#1{\uppercase\expandafter{\romannumeral #1}}
\def\roman#1{\romannumeral #1}
\def\Number#1{\number #1}
\def\unnumberedchapters{\let\makel@bel=\relax \let\chapterlabel=\relax
\let\sectionlabel=\relax \equanumber=-1 }
\def\titlestyle#1{\par\begingroup \interlinepenalty=9999
     \leftskip=0.02\hsize plus 0.23\hsize minus 0.02\hsize
     \rightskip=\leftskip \parfillskip=0pt
     \hyphenpenalty=9000 \exhyphenpenalty=9000
     \tolerance=9999 \pretolerance=9000
     \spaceskip=0.333em \xspaceskip=0.5em
     \iftwelv@\fourteenpoint\else\twelvepoint\fi
   \noindent #1\par\endgroup }
\def\spacecheck#1{\dimen@=\pagegoal\advance\dimen@ by -\pagetotal
   \ifdim\dimen@<#1 \ifdim\dimen@>0pt \vfil\break \fi\fi}
\def\chapter#1{\par \penalty-300 \vskip\chapterskip
   \spacecheck\chapterminspace
   \chapterreset \titlestyle{\chapterlabel \ #1}
   \nobreak\vskip\headskip \penalty 30000
   \wlog{\string\chapter\ \chapterlabel} }

\def\section#1{\par \ifnum\the\lastpenalty=30000\else
   \penalty-200\vskip\sectionskip \spacecheck\sectionminspace\fi
   \wlog{\string\section\ \chapterlabel \the\sectionnumber}
   \global\advance\sectionnumber by 1  \noindent
   {\caps\enspace\chapterlabel \sectionlabel #1}\par
   \nobreak\vskip\headskip \penalty 30000 }
\def\subsection#1{\par
   \ifnum\the\lastpenalty=30000\else \penalty-100\smallskip \fi
   \noindent\undertext{#1}\enspace \vadjust{\penalty5000}}

\def\undertext#1{\vtop{\hbox{#1}\kern 1pt \hrule}}
\def\APPENDIX#1#2{\par\penalty-300\vskip\chapterskip
   \spacecheck\chapterminspace \chapterreset \xdef\chapterlabel{#1}
   \titlestyle{APPENDIX #2} \nobreak\vskip\headskip \penalty 30000
   \wlog{\string\Appendix\ \chapterlabel} }
\def\Appendix#1{\APPENDIX{#1}{#1}}
\def\appendix{\APPENDIX{A}{}}
%
%
%
\def\eqname#1{\relax \ifnum\the\equanumber<0
     \xdef#1{{\rm(\number-\equanumber)}}\global\advance\equanumber by -1
    \else \global\advance\equanumber by 1
      \xdef#1{{\rm(\chapterlabel \number\equanumber)}} \fi}

\def\eqn#1{\eqno\eqname{#1}#1}

\def\eqinsert#1{\noalign{\dimen@=\prevdepth \nointerlineskip
   \setbox0=\hbox to\displaywidth{\hfil #1}
   \vbox to 0pt{\vss\hbox{$\!\box0\!$}\kern-0.5\baselineskip}
   \prevdepth=\dimen@}}
\def\sequentialequations{\globaleqnumbers}
%
%
\def\GENITEM#1;#2{\par \hangafter=0 \hangindent=#1
    \Textindent{$ #2 $}\ignorespaces}
\outer\def\newitem#1=#2;{\gdef#1{\GENITEM #2;}}
\newdimen\itemsize                \itemsize=30pt
\newitem\item=1\itemsize;
\newitem\sitem=1.75\itemsize;     
\newitem\ssitem=2.5\itemsize;     
\outer\def\newlist#1=#2&#3&#4;{\toks0={#2}\toks1={#3}%
   \count255=\escapechar \escapechar=-1
   \alloc@0\list\countdef\insc@unt\listcount     \listcount=0
   \edef#1{\par
      \countdef\listcount=\the\allocationnumber
      \advance\listcount by 1
      \hangafter=0 \hangindent=#4
      \Textindent{\the\toks0{\listcount}\the\toks1}}
   \expandafter\expandafter\expandafter
    \edef\c@t#1{begin}{\par
      \countdef\listcount=\the\allocationnumber \listcount=1
      \hangafter=0 \hangindent=#4
      \Textindent{\the\toks0{\listcount}\the\toks1}}
   \expandafter\expandafter\expandafter
    \edef\c@t#1{con}{\par \hangafter=0 \hangindent=#4 \noindent}
   \escapechar=\count255}
\def\c@t#1#2{\csname\string#1#2\endcsname}
\newlist\point=\Number&.&1.0\itemsize;
\newlist\subpoint=(\alphabetic&)&1.75\itemsize;
\newlist\subsubpoint=(\roman&)&2.5\itemsize;
\newlist\cpoint=\Roman&.&1.0\itemsize;
%

%
%
%
\newcount\referencecount     \referencecount=0
\newif\ifreferenceopen       \newwrite\referencewrite
\newtoks\rw@toks
\def\NPrefmark#1{\atttach{\rm [ #1 ] }}
\let\PRrefmark=\attach
\def\CErefmark#1{\attach{\scriptstyle  #1 ) }}
\def\refmark#1{\relax\ifPhysRev\PRrefmark{#1}\else\NPrefmark{#1}\fi}
\def\crefmark#1{\relax\CErefmark{#1}}
\def\refend{\refmark{\number\referencecount}}
\newcount\lastrefsbegincount \lastrefsbegincount=0
\def\refsend{\refmark{\count255=\referencecount
   \advance\count255 by-\lastrefsbegincount
   \ifcase\count255 \number\referencecount
   \or \number\lastrefsbegincount,\number\referencecount
   \else \number\lastrefsbegincount-\number\referencecount \fi}}
\def\crefsend{\crefmark{\count255=\referencecount
   \advance\count255 by-\lastrefsbegincount
   \ifcase\count255 \number\referencecount
   \or \number\lastrefsbegincount,\number\referencecount
   \else \number\lastrefsbegincount-\number\referencecount \fi}}
\def\refch@ck{\chardef\rw@write=\referencewrite
   \ifreferenceopen \else \referenceopentrue
   \immediate\openout\referencewrite=referenc.texauxil \fi}
%
{\catcode`\^^M=\active 
  \gdef\obeyendofline{\catcode`\^^M\active \let^^M\ }}%
%
{\catcode`\^^M=\active 
  \gdef\ignoreendofline{\catcode`\^^M=5}}
{\obeyendofline\gdef\rw@start#1{\def\t@st{#1} \ifx\t@st\blankend%
\endgroup \@sf \relax \else \ifx\t@st\bl@nkend \endgroup \@sf \relax%
\else \rw@begin#1
\backtotext
\fi \fi } }
{\obeyendofline\gdef\rw@begin#1
{\def\n@xt{#1}\rw@toks={#1}\relax%
\rw@next}}
\def\blankend{}
{\obeylines\gdef\bl@nkend{
}}
\newif\iffirstrefline  \firstreflinetrue
\def\rwr@teswitch{\ifx\n@xt\blankend \let\n@xt=\rw@begin %
 \else\iffirstrefline \global\firstreflinefalse%
\immediate\write\rw@write{\noexpand\obeyendofline \the\rw@toks}%
\let\n@xt=\rw@begin%
      \else\ifx\n@xt\rw@@d \def\n@xt{\immediate\write\rw@write{%
        \noexpand\ignoreendofline}\endgroup \@sf}%
             \else \immediate\write\rw@write{\the\rw@toks}%
             \let\n@xt=\rw@begin\fi\fi \fi}
\def\rw@next{\rwr@teswitch\n@xt}
\def\rw@@d{\backtotext} \let\rw@end=\relax
\let\backtotext=\relax

\newdimen\refindent     \refindent=30pt
\def\refitem#1{\par \hangafter=0 \hangindent=\refindent \Textindent{#1}}
\def\REFNUM#1{\space@ver{}\refch@ck \firstreflinetrue%
 \global\advance\referencecount by 1 \xdef#1{\the\referencecount}}
\def\refnum#1{\space@ver{}\refch@ck \firstreflinetrue%
 \global\advance\referencecount by 1 \xdef#1{\the\referencecount}\refend}

\def\REF#1{\REFNUM#1%
 \immediate\write\referencewrite{%
 \noexpand\refitem{#1.}}%
\begingroup\obeyendofline\rw@start}
\def\ref{\refnum\?%
 \immediate\write\referencewrite{\noexpand\refitem{\?.}}%
\begingroup\obeyendofline\rw@start}
\def\Ref#1{\refnum#1%
 \immediate\write\referencewrite{\noexpand\refitem{#1.}}%
\begingroup\obeyendofline\rw@start}
\def\REFS#1{\REFNUM#1\global\lastrefsbegincount=\referencecount
\immediate\write\referencewrite{\noexpand\refitem{#1.}}%
\begingroup\obeyendofline\rw@start}
\newcount\figurecount     \figurecount=0
\newif\iffigureopen       \newwrite\figurewrite
\def\figch@ck{\chardef\rw@write=\figurewrite \iffigureopen\else
   \immediate\openout\figurewrite=figures.texauxil
   \figureopentrue\fi}
\def\FIGNUM#1{\space@ver{}\figch@ck \firstreflinetrue%
 \global\advance\figurecount by 1 \xdef#1{\the\figurecount}}
\def\FIG#1{\FIGNUM#1
   \immediate\write\figurewrite{\noexpand\refitem{#1.}}%
   \begingroup\obeyendofline\rw@start}
\def\fig{\FIGNUM\? fig.~\?%
\immediate\write\figurewrite{\noexpand\refitem{\?.}}%
\begingroup\obeyendofline\rw@start}
\def\figure{\FIGNUM\? figure~\?
   \immediate\write\figurewrite{\noexpand\refitem{\?.}}%
   \begingroup\obeyendofline\rw@start}
\def\Fig{\FIGNUM\? Fig.~\?%
\immediate\write\figurewrite{\noexpand\refitem{\?.}}%
\begingroup\obeyendofline\rw@start}
\def\Figure{\FIGNUM\? Figure~\?%
\immediate\write\figurewrite{\noexpand\refitem{\?.}}%
\begingroup\obeyendofline\rw@start}
\newcount\tablecount     \tablecount=0
\newif\iftableopen       \newwrite\tablewrite
\def\tabch@ck{\chardef\rw@write=\tablewrite \iftableopen\else
   \immediate\openout\tablewrite=tables.texauxil
   \tableopentrue\fi}
\def\TABNUM#1{\space@ver{}\tabch@ck \firstreflinetrue%
 \global\advance\tablecount by 1 \xdef#1{\the\tablecount}}
\def\TABLE#1{\TABNUM#1
   \immediate\write\tablewrite{\noexpand\refitem{#1.}}%
   \begingroup\obeyendofline\rw@start}
\def\Table{\TABNUM\? Table~\?%
\immediate\write\tablewrite{\noexpand\refitem{\?.}}%
\begingroup\obeyendofline\rw@start}
%

%
%
%
\def\masterreset{\global\pagenumber=1 \global\chapternumber=0
   \ifnum\the\equanumber<0\else \global\equanumber=0\fi
   \global\sectionnumber=0
   \global\referencecount=0 \global\figurecount=0 \global\tablecount=0 }
\def\FRONTPAGE{\ifvoid255\else\vfill\penalty-2000\fi
      \masterreset\global\frontpagetrue
      \global\lastf@@t=0 \global\footsymbolcount=0}

\def\paperstyle{\letterstylefalse\normalspace\papersize}
\def\letterstyle{\letterstyletrue\singlespace\lettersize}
\def\papersize{\hsize=6.5truein\vsize=9.1truein\hoffset=-.3truein
               \voffset=-.4truein\skip\footins=\bigskipamount}
\def\lettersize{\hsize=6.5truein\vsize=9.1truein\hoffset=-.3truein
    \voffset=.1truein\skip\footins=\smallskipamount \multiply
    \skip\footins by 3 }
\paperstyle   
%
%
\def\MEMO{\letterstyle\FRONTPAGE \letterfrontheadline={\hfil}
    \line{\quad\fourteenrm CERN MEMORANDUM\hfil\twelverm\the\date\quad}
    \medskip \memod@f}

\def\memit@m#1{\smallskip \hangafter=0 \hangindent=1in
      \Textindent{\caps #1}}
\def\memod@f{\xdef\mto{\memit@m{To:}}\xdef\from{\memit@m{From:}}%
     \xdef\topic{\memit@m{Topic:}}\xdef\subject{\memit@m{Subject:}}%
     \xdef\rule{\bigskip\hrule height 1pt\bigskip}}
\memod@f
\newskip\lettertopfil
\lettertopfil = 0pt plus 1.5in minus 0pt
\newskip\letterbottomfil
\letterbottomfil = 0pt plus 2.3in minus 0pt
\newskip\spskip \setbox0\hbox{\ } \spskip=-1\wd0
\def\addressee#1{\medskip\rightline{\the\date\hskip 30pt} \bigskip
   \vskip\lettertopfil
   \ialign to\hsize{\strut ##\hfil\tabskip 0pt plus \hsize \cr #1\crcr}
   \medskip\noindent\hskip\spskip}
\newskip\signatureskip       \signatureskip=40pt
\def\signed#1{\par \penalty 9000 \bigskip \dt@pfalse
  \everycr={\noalign{\ifdt@p\vskip\signatureskip\global\dt@pfalse\fi}}
  \setbox0=\vbox{\singlespace \halign{\tabskip 0pt \strut ##\hfil\cr
   \noalign{\global\dt@ptrue}#1\crcr}}
  \line{\hskip 0.5\hsize minus 0.5\hsize \box0\hfil} \medskip }

\def\endletter{\ifnum\pagenumber=1 \vskip\letterbottomfil\supereject
\else \vfil\supereject \fi}
\newbox\letterb@x
\def\lettertext{\par\unvcopy\letterb@x\par}
\def\multiletter{\setbox\letterb@x=\vbox\bgroup
      \everypar{\vrule height 1\baselineskip depth 0pt width 0pt }
      \singlespace \topskip=\baselineskip }
\def\letterend{\par\egroup}
%
%
%
\newskip\frontpageskip
\newtoks\pubtype
\newtoks\Pubnum
\newtoks\pubnum
\newtoks\pubnu
\newtoks\pubn
\newif\ifp@bblock  \p@bblocktrue
\def\PH@SR@V{\doubl@true \baselineskip=24.1pt plus 0.2pt minus 0.1pt
             \parskip= 3pt plus 2pt minus 1pt }
\def\PHYSREV{\paperstyle\PhysRevtrue\PH@SR@V}
\def\titlepage{\FRONTPAGE\paperstyle\ifPhysRev\PH@SR@V\fi
   \ifp@bblock\p@bblock\fi}
\def\nopubblock{\p@bblockfalse}
\def\endpage{\vfil\break}
\frontpageskip=1\medskipamount plus .5fil
\pubtype={\tensl Preliminary Version}
\Pubnum={$\rm CERN-TH.\the\pubnum $}
\pubnum={0000}
\def\p@bblock{\begingroup \tabskip=\hsize minus \hsize
   \baselineskip=1.5\ht\strutbox \topspace-2\baselineskip
   \halign to\hsize{\strut ##\hfil\tabskip=0pt\crcr
   \the \pubn\cr
   \the \Pubnum\cr
   \the \pubnu\cr
   \the \date\cr}\endgroup}
\def\title#1{\vskip\frontpageskip \titlestyle{#1} \vskip\headskip }
\def\author#1{\vskip\frontpageskip\titlestyle{\twelvecp #1}\nobreak}

\def\address#1{\par\kern 5pt\titlestyle{\twelvepoint\it #1}}
\def\andaddress{\par\kern 5pt \centerline{\sl and} \address}

\def\abstract{\vskip\frontpageskip\centerline{\fourteenrm ABSTRACT}
              \vskip\headskip }

%
%
%

\def\\{\relax\ifmmode\backslash\else$\backslash$\fi}
\def\globaleqnumbers{\relax\ifnum\the\equanumber<0%
\else\global\equanumber=-1\fi}
\def\nextline{\unskip\nobreak\hskip\parfillskip\break}

\def\journal#1&#2(#3){\unskip, \sl #1~\bf #2 \rm (19#3) }

\def\topspace{\hrule height 0pt depth 0pt \vskip}

\let\int=\intop         
\def\prop{\mathrel{{\mathchoice{\pr@p\scriptstyle}{\pr@p\scriptstyle}{
                \pr@p\scriptscriptstyle}{\pr@p\scriptscriptstyle} }}}
\def\pr@p#1{\setbox0=\hbox{$\cal #1 \char'103$}
   \hbox{$\cal #1 \char'117$\kern-.4\wd0\box0}}
\def\lsim{\mathrel{\mathpalette\@versim<}}
\def\gsim{\mathrel{\mathpalette\@versim>}}
\def\@versim#1#2{\lower0.2ex\vbox{\baselineskip\z@skip\lineskip\z@skip
  \lineskiplimit\z@\ialign{$\m@th#1\hfil##\hfil$\crcr#2\crcr\sim\crcr}}}
\def\leftrightarrowfill{$\m@th \mathord- \mkern-6mu
        \cleaders\hbox{$\mkern-2mu \mathord- \mkern-2mu$}\hfil
        \mkern-6mu \mathord\leftrightarrow$}
\def\lrover#1{\vbox{\ialign{##\crcr
        \leftrightarrowfill\crcr\noalign{\kern-1pt\nointerlineskip}
        $\hfil\displaystyle{#1}\hfil$\crcr}}}
%
%
%
\let\sec@nt=\sec
\def\sec{\relax\ifmmode\let\n@xt=\sec@nt\else\let\n@xt\section\fi\n@xt}
\def\obsolete#1{\message{Macro \string #1 is obsolete.}}
\def\firstsec#1{\obsolete\firstsec \section{#1}}
\def\firstsubsec#1{\obsolete\firstsubsec \subsection{#1}}
\def\thispage#1{\obsolete\thispage \global\pagenumber=#1\frontpagefalse}
\def\thischapter#1{\obsolete\thischapter \global\chapternumber=#1}
\def\nextequation#1{\obsolete\nextequation \global\equanumber=#1
   \ifnum\the\equanumber>0 \global\advance\equanumber by 1 \fi}
\def\BOXITEM{\afterassigment\B@XITEM\setbox0=}
\def\B@XITEM{\par\hangindent\wd0 \noindent\box0 }
%

%
%

%
%

%
%

%

%

%

%

%

%
%
%
\def\boxit#1{\vbox{\hrule\hbox{\vrule\kern3pt\vbox{\kern3pt#1\kern3pt}
\kern3pt\vrule}\hrule}}
%
%
%
\catcode`@=12 
\message{ by V.K./U.B.}
\everyjob{\input imyphyx }
%
%
%
%
%
%
%
%
%
%
%
%
%
\catcode`@=11

\font\seventeencp=cmcsc10 scaled\magstep3
\def\SIZE{\hsize=6.6truein\vsize=9.1truein}
\def\OFFSET{\voffset=1.2truein\hoffset=.8truein}
\def\papersize{\SIZE\OFFSET\skip\footins=\bigskipamount
\normaldisplayskip= 30pt plus 5pt minus 10pt}
\Pubnum={\rm CERN$-$TH.\the\pubnum }
\def\title#1{\vskip\frontpageskip\vskip .50truein
     \titlestyle{\seventeencp #1} \vskip\headskip\vskip\frontpageskip
     \vskip .2truein}
\def\author#1{\vskip .27truein\titlestyle{#1}\nobreak}

\def\p@bblock{\begingroup \tabskip=\hsize minus \hsize
   \baselineskip=1.5\ht\strutbox \topspace-2\baselineskip
   \halign to\hsize{\strut ##\hfil\tabskip=0pt\crcr
   \the \Pubnum\cr}\endgroup}
\def\makefootline{\iffrontpage\vskip .27truein\line{\the\footline}
                 \vskip -.1truein\line{\the\date\hfil}
              \else\line{\the\footline}\fi}
\paperfootline={\iffrontpage \the\Pubnum\hfil\else\hfil\fi}
\paperheadline={\iffrontpage\hfil
                \else\twelverm\hss $-$\ \folio\ $-$\hss\fi}
\newif\ifmref  
\newif\iffref  
\def\xrefsend{\xrefmark{\count255=\referencecount
\advance\count255 by-\lastrefsbegincount
\ifcase\count255 \number\referencecount
\or \number\lastrefsbegincount,\number\referencecount
\else \number\lastrefsbegincount-\number\referencecount \fi}}
\def\xrefsdub{\xrefmark{\count255=\referencecount
\advance\count255 by-\lastrefsbegincount
\ifcase\count255 \number\referencecount
\or \number\lastrefsbegincount,\number\referencecount
\else \number\lastrefsbegincount,\number\referencecount \fi}}
\def\xREFNUM#1{\space@ver{}\refch@ck\firstreflinetrue%
\global\advance\referencecount by 1
\xdef#1{\xrefend}}
\def\xrefend{\xrefmark{\number\referencecount}}
\def\xrefmark#1{[{#1}]}
\def\xRef#1{\xREFNUM#1\immediate\write\referencewrite%
{\noexpand\refitem{#1}}\begingroup\obeyendofline\rw@start}%
\def\xREFS#1{\xREFNUM#1\global\lastrefsbegincount=\referencecount%
\immediate\write\referencewrite{\noexpand\refitem{#1}}%
\begingroup\obeyendofline\rw@start}
\def\rrr#1#2{\relax\ifmref{\iffref\xREFS#1{#2}%
\else\xRef#1{#2}\fi}\else\xRef#1{#2}\xrefend\fi}
\referencecount=0
%
\space@ver{}\refch@ck\firstreflinetrue%
\immediate\write\referencewrite{}%
\begingroup\obeyendofline\rw@start{}%
\def\plb#1({Phys.\ Lett.\ $\underline  {#1B}$\ (}
\def\nup#1({Nucl.\ Phys.\ $\underline {B#1}$\ (}
\def\plt#1({Phys.\ Lett.\ $\underline  {B#1}$\ (}
\def\cmp#1({Comm.\ Math.\ Phys.\ $\underline  {#1}$\ (}
\def\prp#1({Phys.\ Rep.\ $\underline  {#1}$\ (}
\def\prl#1({Phys.\ Rev.\ Lett.\ $\underline  {#1}$\ (}
\def\prv#1({Phys.\ Rev. $\underline  {D#1}$\ (}
\def\und#1({            $\underline  {#1}$\ (}
\message{ by W.L.}
\everyjob{\input offset }
\catcode`@=12

\let\it=\sl

\def\OFFSET{\hoffset=6.pt\voffset=40.pt}
\def\SIZE{\hsize=420.pt\vsize=620.pt}
\OFFSET
\def\PLANCK{\rrr\PLANCK{L.E. Ib\'a\~nez, \plb126 (1983) 196;
\nextline J.E. Bjorkman and D.R.T. Jones, \nup259 (1985) 533.}}

\def\ZNZM{\rrr\ZNZM{A. Font, L.E. Ib\'a\~nez and F. Quevedo,
\plb217 (1989) 272.}}

\def\IN{\rrr\IN{L.E. Ib\'a\~nez and H.P. Nilles, \plb169 (1986) 354.}}

\def\EINHJON{\rrr\EINHJON{M. Einhorn and D.R.T. Jones, \nup196 (1982)
                      475.}}

\def\KAPLU{\rrr\KAPLU{V. Kaplunovsky, \nup307 (1988) 145.}}

\def\AMALDI{\rrr\AMALDI{
J. Ellis, S. Kelley and D.V. Nanopoulos, \plb249 (1990) 441;
\plb260 (1991) 131;
P. Langacker, {\it ``Precision tests of the standard model",}
Pennsylvania preprint UPR-0435T, (1990);
U. Amaldi, W. de Boer and H. F\"urstenau, \plt260 (1991) 447;
P. Langacker and M. Luo, Phys.Rev.D44 (1991) 817;
R. Roberts and G.G. Ross, talk presented by G.G. Ross at 1991 Joint
International Lepton-Photon Symposium and EPS Conference, and to
be published.
}}

\def\DG{\rrr\DG{S. Dimopoulos, S. Raby and F. Wilczek, Phys. Rev.
D24 (1981) 1681;\nextline
                 L.E. Ib\'a\~nez and G.G. Ross, \plb105 (1981) 439;
\nextline S. Dimopoulos and H. Georgi, \nup193 (1981) 375.}}

\def\IMNQ{\rrr\IMNQ{L.E. Ib\'a\~nez, H.P. Nilles and F. Quevedo,
\plt187 (1987) 25; L.E. Ib\'a\~nez, J. Mas, H.P. Nilles and
F. Quevedo, \nup301 (1988) 157; A. Font, L.E. Ib\'a\~nez,
F. Quevedo and A. Sierra, \nup331 (1990) 421.}}

\def\SCHELL{\rrr\SCHELL{A.N. Schellekens, \plt237 (1990) 363.}}

\def\GQW{\rrr\GQW{H. Georgi, H.R. Quinn and S. Weinberg, Phys. Rev.
Lett. ${\underline{33}}$ (1974) 451.}}

\def\GINS{\rrr\GINS{P. Ginsparg, \plt197 (1987) 139.}}

\def\ELLISETAL{\rrr\ELLISETAL{I. Antoniadis, J. Ellis, R. Lacaze
and D.V. Nanopoulos, {\it ``String Threshold Corrections and
                Flipped $SU(5)$'',} preprint CERN-TH.6136/91 (1991);
    S. Kalara, J.L. Lopez and D.V. Nanopoulos, {\it``Threshold
   Corrections and Modular Invariance in Free Fermionic Strings'',}
      preprint CERN-TH-6168/91 (1991).}}

\def\DFKZ{\rrr\DFKZ{J.P. Derendinger, S. Ferrara, C. Kounnas and
F. Zwirner,{\it ``On loop corrections to string effective field theories:
         field-dependent gauge couplings and sigma-model anomalies'',}
        preprint CERN-TH.6004/91, LPTENS 91-4 (revised version) (1991).}}

\def\LOUIS{\rrr\LOUIS{J. Louis, {\it
         ``Non-harmonic gauge coupling constants in supersymmetry
         and superstring theory'',} preprint SLAC-PUB-5527 (1991);
         V. Kaplunovsky and J. Louis, as quoted in J. Louis,
         SLAC-PUB-5527 (1991).}}

\def\DIN{\rrr\DIN{J.P. Derendinger, L.E. Ib\'a\~nez and H.P Nilles,
        \nup267 (1986) 365.}}

\def\DHVW{\rrr\DHVW{L. Dixon, J. Harvey, C.~Vafa and E.~Witten,
         \nup261 (1985) 651;
        \nup274 (1986) 285.}}

\def\DKLB{\rrr\DKLB{L. Dixon, V. Kaplunovsky and J. Louis,
         \nup355 (1991) 649.}}

\def\DKLA{\rrr\DKLA{L. Dixon, V. Kaplunovsky and J. Louis, \nup329 (1990)
            27.}}

\def\ALOS{\rrr\ALOS{E. Alvarez and M.A.R. Osorio, \prv40 (1989) 1150.}}

\def\FILQ{\rrr\FILQ{A. Font, L.E. Ib\'a\~nez, D. L\"ust and F. Quevedo,
           \plt245 (1990) 401.}}

\def\CFILQ{\rrr\CFILQ{M. Cvetic, A. Font, L.E.
           Ib\'a\~nez, D. L\"ust and F. Quevedo, \nup361 (1991) 194.}}

\def\FILQ{\rrr\FILQ{A. Font, L.E. Ib\'a\~nez, D. L\"ust and F. Quevedo,
           \plt245 (1990) 401.}}

\def\DUAGAU{\rrr\DUAGAU{S. Ferrara, N. Magnoli, T.R. Taylor and
           G. Veneziano, \plt245 (1990) 409; H.P. Nilles and M.
           Olechowski, \plt248 (1990) 268; P. Binetruy and M.K.
           Gaillard, \plt253 (1991) 119; J. Louis, SLAC-PUB-5645
           (1991); S. Kalara, J. Lopez and D. Nanopoulos,
           Texas A\&M  preprint CTP-TAMU-69/91.}}

\def\CFILQ{\rrr\CFILQ{M. Cvetic, A. Font, L.E.
           Ib\'a\~nez, D. L\"ust and F. Quevedo, \nup361 (1991) 194.}}

\def\FIQ{\rrr\FIQ{A. Font, L.E. Ib\'a\~nez and F. Quevedo,
        \plt217 (1989) 272.}}

\def\FLST{\rrr\FLST{S. Ferrara,
         D. L\"ust, A. Shapere and S. Theisen, \plt225 (1989) 363.}}

\def\FLT{\rrr\FLT
{S. Ferrara, D. L\"ust and S. Theisen, \plt233 (1989) 147.}}

\def\IBLU{\rrr\IBLU{L. Ib\'a\~nez and D. L\"ust,
          \plb267 (1991) 51.}}

\def\GAUGINO{\rrr\GAUGINO{J.P. Derendinger, L.E. Ib\'a\~nez and H.P. Nilles,
            \plb155 (1985) 65;
         M. Dine, R. Rohm, N. Seiberg and E. Witten, \plb156 (1985) 55.}}

\def\GHMR{\rrr\GHMR{D.J. Gross, J.A. Harvey, E. Martinec and R. Rohm,
         \prl54 (1985) 502; \nup256 (1985) 253; \nup267 (1986) 75.}}

\def\ANT{\rrr\ANT{I. Antoniadis, K.S. Narain and T.R. Taylor,
        \plb267 (1991) 37.}}

\def\WEIN{\rrr\WEIN{S. Weinberg, \plb91 (1980) 51.}}

\def\WITTEF{\rrr\WITTEF{E. Witten, \plb155 (1985) 151.}}

\def\HB{\rrr\HB{L. Hall and R. Barbieri, private communication and
preprint in preparation (1991).}}

\def\SCHELL{\rrr\SCHELL{A.N. Schellekens, ``Superstring
construction'', North Holland, Amsterdam (1989).}}

\def\FLIP{\rrr\FLIP{I. Antoniadis, J. Ellis, J. Hagelin and
D.V. Nanopoulos, \plb231 (1989) 65 and references therein. For a
recent review see J. Lopez and D.V. Nanopoulos,
                  Texas A\& M preprint CTP-TAMU-76/91 (1991).}}

\def\GKMR{\rrr\GKPM{B. Greene, K. Kirklin, P. Miron and G.G. Ross,
\nup292 (1987) 606.}}

\def\OTH{\rrr\OTH{D. Bailin, A. Love and S. Thomas, \plb188 (1987)
193; \plb194 (1987) 385; B. Nilsson, P. Roberts and P. Salomonson,
\plb222 (1989) 35;
J.A. Casas, E.K. Katehou and C. Mu\~noz, \nup317 (1989) 171;
J.A. Casas and C. Mu\~noz, \plb209 (1988) 214, \plb212 (1988) 343
J.A. Casas, F. Gomez and C. Mu\~noz, \plb251 (1990) 99;
A. Chamseddine and J.P. Derendinger, \nup301 (1988) 381;
A. Chamseddine and M. Quiros, \plb212 (1988) 343, \nup316 (1989) 101;
T. Burwick, R. Kaiser and H. M\"uller, \nup355 (1991) 689;
Y. Katsuki, Y. Kawamura, T. Kobayashi, N. Ohtsubo,
Y. Ono and K. Tanioka, \nup341 (1990) 611.}}

\def\SCHLUS{\rrr\SCHLUS{For a review, see e.g. J. Schwarz, Caltech
preprint CALT-68-1740 (1991); D. L\"ust,  CERN preprint TH.6143/91.
}}

\def\LMN{\rrr\LMN{J. Lauer, J. Mas and H.P. Nilles, \plb226 (1989)
251, \nup351 (1991) 353; W. Lerche, D. L\"ust and N.P. Warner,
\plb231 (1989) 417.}}

\def\ILR{\rrr\ILR{ L.E. Ib\'a\~nez, D. L\"ust and G.G. Ross,
CERN preprint TH.6241/91 (1991).}}

\def\ANTON{\rrr\ANTON{I. Antoniadis, J. Ellis, S. Kelley and
D.V. Nanopoulos, CERN preprint TH.6169/91 (1991).}}

\def\HIGH{\rrr\HIGH{D. Lewellen, \nup337 (1990) 61;
J.A. Schwartz, Phys.Rev. D42 (1990) 1777.}}

\def\HIGHK{\rrr\HIGHK{A. Font, L.E. Ib\'a\~nez and F. Quevedo,
\nup345 (1990) 389; J. Ellis, J. Lopez and D.V. Nanopoulos,
\plb245 (1990) 375.}}

\def\LLR{\rrr\LLR{C.H. Llewellyn-Smith, G.G. Ross and
J.F. Wheater, \nup177 (1981) 263.}}

\def\IL{\rrr\IL{  L.E. Ib\'a\~nez and D. L\"ust, CERN preprint (1991)
to appear.}}

\def\YO{\rrr\YO{ L.E. Ib\'a\~nez, {\it``Some topics
in the low energy physics from superstrings''} in proceedings of
the NATO workshop on ``Superfield Theories", Vancouver,
Canada. Plenum Press, New York (1987).}}

\def\SDUAL{\rrr\SDUAL{A. Font, L.E. Ib\'a\~nez, D. L\"ust
and F. Quevedo, \plb249 (1990) 35.}}

\def\MALLOR{\rrr\MALLOR{For a recent review see L.E. Ib\'a\~nez,
{\it ``Beyond the Standard Model (yet again)''}, CERN preprint
TH.5982/91, to appear in the Proceedings of the 1990 CERN
School of Physics, Mallorca (1990).}}

\def\FIQS{\rrr\FIQS{A. Font, L.E. Ib\'a\~nez, F. Quevedo and
A. Sierra, \nup337 (1990) 119.}}

\def\LT{\rrr\LT{D. L\"ust and T.R. Taylor, \plb253 (1991) 335;
B. Carlos, J. Casas and C. Mu\~noz, preprint CERN-TH.6049/91
(1991).}}

\def\OFFSET{\hoffset=12.pt\voffset=55.pt}
\def\SIZE{\hsize=420.pt\vsize=620.pt}

\catcode`@=12
\newtoks\Pubnumtwo
\newtoks\Pubnumthree
\catcode`@=11
\def\p@bblock{\begingroup\tabskip=\hsize minus\hsize
   \baselineskip=0.5\ht\strutbox\topspace-2\baselineskip
   \halign to \hsize{\strut ##\hfil\tabskip=0pt\crcr
   \the\Pubnum\cr  \the\Pubnumtwo\cr 
   \the\pubtype\cr}\endgroup}
\pubnum={6342/91}
\date{December  1991}
\pubtype={}
\titlepage
\vskip -.6truein
\title{
Topics in String Unification}
 \centerline{\bf Luis E. Ib\'a\~nez}
 \vskip .1truein
   \vskip 0.1truein
  \centerline{CERN, 1211 Geneva 23, Switzerland}
\vskip 0.1truein
\abstract\noindent\nobreak
I discuss several aspects of strings as unified theories. After
recalling the difficulties of the simplest supersymmetric grand
unification schemes I emphasize the distinct features of string
unification. An important role in constraining the effective low
energy physics from strings is played by $duality$ symmetries.
The discussed topics include the unification of coupling constants
(computation of $\sin ^2\theta _W$ and $\alpha _s$ at the weak scale),
supersymmetry breaking through gaugino condensation, and properties
of the induced SUSY-breaking soft terms. I remark that
departures from universality in the soft terms are (in contrast
to the minimal SUSY     model) generically expected.

\vskip 1.0cm

\bigskip
Talk given at the ``Workshop on Electroweak Physics Beyond the
Standard Model'', Valencia, 3-5 October 1991.
\endpage
\pagenumber=1
\sequentialequations

\leftline{\bf 1. Strings as unified theories}
\bigskip

The recent LEP precision  measurements of the Standard Model (SM)
gauge-coupling parameters have confirmed \AMALDI \ the remarkable
agreement with the expectations from supersymmetric grand
unification \DG   , \EINHJON   . If taken seriously, this agreement would
suggest the existence of a supersymmetric GUT like $SU(5)$ or $SO(10)$
            beyond an energy scale $M_{X}\equiv 10^{16}$ GeV.

While the prediction for $\sin ^2\theta _W$ in these models
works remarkably well, one must recall, however, that these
theories have several important theoretical problems.
Perhaps the more severe one is that of understanding the
huge mass splitting between the Higgs doublets of the
theory and the colour triplet chiral fields which come
along in any GUT scheme (the $doublet-triplet$ $splitting$ problem).
It is not just that we do not understand why this splitting
occurs; the worst  problem is that the          mechanisms
considered up to now $destroy $ $the$ $hierarchy$ of masses, i.e.
give, either at tree level or in loops, a huge mass to the
standard Higgs doublets. The simplest example         of this fact is
just the minimal SUSY-$SU(5)$ model with SUSY-breaking soft
terms. There you get doublet-triplet splitting by fine-tuning
of the couplings $\lambda $ and $\mu $ in the superpotential
$\lambda \phi _{24}H_5H_{\bar 5}\ +\ \mu H_5H_{\bar 5}$.
However once SUSY is broken you get soft SUSY-breaking couplings
$Am\lambda \phi _{24}H_5H_{\bar 5}\ +\ Bm\mu H_5H_{\bar 5}$, where
$A,B,m$ are independent SUSY-breaking parameters. Now the fine-tuning
          that  guaranteed the masslessness of the Higgs doublets
in the superpotential is spoiled by the contribution of the
soft terms which, for generic $A,B$, gives a huge mass to the
Higgses . This is not just a problem of the minimal model but
appears to be quite generic in SUSY-GUTs, although it sometimes
appears only at one loop          or two loops \HB        .
The origin of the problem is that soft susy-breaking terms
spoil the hierarchy whenever there are fields that couple
both to the doublets and other supermassive fields.
Apart from this generic problem, it is also well known that usual GUTs
have difficulties with   their predictions for quark-lepton masses
of the first two generations. Furthermore, the breaking
of the GUT symmetry generally requires huge Higgs representations,
especially if one wants to get consistent quark-lepton mass spectra.

In view of the problems reviewed above, one should take the
success of standard SUSY-GUTs with a grain of salt. There are
further theoretical reasons to believe that such schemes
are too naive to be true. To start with, the standard
coupling of these theories to $N=1$ supergravity (required
in order to obtain the appropriate soft terms in a natural way)
leads to a non-renormalizable theory. Although below the Planck
mass one can deal with a renormalizable effective Lagrangian,
eventually one will have to face the problem of
a non-renormalizable quantum gravity.

Four-dimensional supersymmetric strings are obvious candidates
for unified theories since they provide us ,in principle,
with the first finite theories of all interactions, including
gravity. However, string theory is $not$ a model for
unification, it is an alternative to field theory itself.
Below the string scale ($\simeq M_{Planck} $) one
expects to describe the low-energy physics in terms
of some standard field theory, presumably  some sort
of $N=1$ supergravity Lagrangian coupled to the
standard model (or some of its gauge extensions).
The actual challenge is trying to see what constraints
(if any) should obey such a Lagrangian because of its $stringy$
$origin$. Asking for a model-independent test of string
theory is perhaps too much in the same sense as   asking for
a model-independent test of $field$ theory (without any reference to
the actual interactions realized in nature) is itself hard!
But one can easily see that the building of string versions
of the standard model or of possible extensions  is quite
constrained:     string model building has its own rules.
Symmetries such as superconformal and modular invariance on the
   worldsheet     as well as target-space duality symmetries
substantially constrain  the particle content and interactions
of the string models.

There are several techniques to build 4-D strings, and typically a
given model may be constructed in several different ways \SCHELL .
Several examples of 4-D string theories with three generations
of quarks and leptons (plus extra exotics) exist in the literature
\IMNQ ,\GKMR ,\FLIP ,\OTH .
Very often (maybe always \FIQS )  one can derive the four-dimensional
string starting from the 10-D  heterotic string and compactifying
the 6 extra dimensions into an unobservable space with overall
size $R\equiv 1/10^{18}\ GeV^{-1}$. The massless sector of the
theory contains the standard pure $N=1$ supergravity particles,
plus gauge particles and matter fields. There are essentially
three types of massless matter fields in the massless sector:
i) Charged matter fields $\phi _{\alpha }$. These include
the standard quark, lepton, and Higgs superfields.
ii) The $moduli$ fields $T_i$. These are massless fields, which are
singlets under the observed gauge interactions. Some of them
are associated to the size and shape of the compactifying
manifold. In particular, the real part of one of them (usually denoted
$T$ ) is related to the compactification radius ($ReT=R^2$).
iii) The $dilaton$ field $S$. This is a singlet field whose
real part is related to the tree-level gauge coupling constant
($ReS=1/g^2$).
One of the specific features of 4-D strings is that
coupling constants are in fact fields whose vacuum
expectation values correspond to the measured values.
The couplings depend on the  dilaton  field $S$ as well
as on the moduli fields  $T_i$. Unfortunately  these fields
have a vanishing scalar potential in perturbation theory
and hence we need a certain knowledge of the non-perturbative
effects which may eventually determine $<S>$ and $<T_i>$.
However, we may still get relationships amongst different
coupling constants assuming that eventually the dynamics
will fix those vev's (some recent ideas about how this
may happen are discussed below).

One important point to remark is that, $in$ $string$
$theory$, $gauge$ $interactions$ $are$ $necessarily$
$unified$ even in the absence of a unification gauge
group such as $SU(5)$. Thus, for instance, if one builds an
$SU(3)\times SU(2)\times U(1)$  string, the gauge
coupling constants are unified with the Newton
coupling constant as follows \GINS :

$$g_1^2k_1 \ =\ g_2^2k_2\ =\ g_3^2k_3\ =\ {{4\pi }\over {\alpha '}}
G_{Newton} ,\eqn \unif $$
where $k_i$ are the Kac-Moody levels
                             of the $U(1)$, $SU(2)$ and $SU(3)$
factors and $\alpha '$ is the inverse of the string tension
squared.
        For the case of $SU(2)$ and $SU(3)$, the levels are
integer numbers. For the case of $U(1)$, the level $k_1$
is just a normalization factor of the hypercharge, and it is
a  rational number in the specific models constructed up to now.
Most of the string models constructed up to now have
non-Abelian Kac-Moody levels $=1$. Concerning $k_1$, many
models have in fact $k_1 >   5/3$.
The boundary condition of standard GUTs is obtained for
$k_3=k_2=3/5k_1$, but other possibilities are in general
allowed in a non-unified $SU(3)\times SU(2)\times U(1)$
string model.

\endpage

\leftline{\bf 2. Duality symmetries}

It has been realized in the last three years that
in generic 4-D strings the spectrum and interactions
are invariant under certain discrete infinite groups
called duality symmetries \SCHLUS . Those are transformations
in the space of the moduli $T_i$ which also induce
$T_i$-dependent transformations in the rest of the
massless chiral fields. In the case of the overall
modulus $T\equiv R^2+i\eta $, the discrete group has two types
of generators: i) One that  relates small and large $R^2$
and ii) a sort of discrete Peccei-Quinn symmetry under
which $T\rightarrow T+i$. The first symmetry is remarkable
because it tells us that there is a minimum physical scale
below which the physics will be just equivalent. The
prototype duality symmetry is $target$ $space$ $modular$
$invariance$ which is present in orbifold-like 4-dimensional
strings. This symmetry acting on the overall modulus $T$
is given     by the class of transformations
$$\eqalign{
T\ \longrightarrow {{a\ T\ -\ ib}\over {ic\ T\ +\ d}}
\ \ ;\ \ \ & a,b,c,d\in {\bf Z},\cr
\ & ad-bc=1 \ .\cr}
\eqn \mod $$
This corresponds to the discrete infinite group $SL(2,{\bf Z})$
generated by $T\rightarrow 1/T$ and $T\rightarrow T+i$.
It has been shown that this is a symmetry of the string
spectrum and interactions order by order in perturbation
theory \ALOS \ on the string coupling constant ($S$). In what follows
we are going to concentrate on this specific example
of duality symmetry, which is the one relevant for one
of the largest known classes of 4-D strings, that of
Abelian $Z_N$ \DHVW ,\IMNQ \ and $Z_N\times Z_M$ \ZNZM \ orbifolds.

Since the string theory is invariant under this symmetry,
the effective low-energy Lagrangian will also be invariant
under it. In the case of an $N=1$ supersymmetric model, the
Lagrangian will be determined by three functions:
the Kahler potential $K(\phi _{\alpha },T_i,S)$, the
superpotential $W$, and the gauge kinetic function $f$.
Considering, for simplicity, the dependence on the
overall modulus $T$, it is well known that the
tree-level Kahler potential for orbifold-type 4-D
strings has the form \WITTEF ,\DKLA ,\FLT :
$$
K(\phi _{\alpha },T,S)\ =\ -\log (S+S^{*} )\ -\ 3\log (T+T^{*})
\ +\ \sum _j\ (T+T^{*})^{n_j}\ \phi _j^{*}\phi _j  \eqn \kahl
$$
to first order in the $\phi $'s. The $n_j$'s are integers
($n_j=-1,-2$ for untwisted and twisted matter fields.
Those values are increased or  decreased  in units depending on the
number of twisted oscillators involved in the vertex \LOUIS ,
\IL ).                          Since the $S,T$ fields have
no superpotential (order by order in perturbation theory), one can
see that the complete superpotential will be a  holomorphic
function of $\phi _{\alpha }$ and $T$, $W(\phi _{\alpha },T)$.
The complete Lagrangian depends on $K$ and $W$ only through
the combination
$$
G\ \equiv \ K \ +\ \log |W|^2 \ .\eqn \ggg
$$
One can check that $G$ is invariant under the modular
transformation  \mod\ provided the matter fields $\phi _j        $
and the superpotential $W(\phi _j        , T)$ transform
like
$$
\eqalign{ \phi _j\ \longrightarrow &\ \phi _j\ (icT\ +\ d)^{n_j}  \cr
W(\phi _j,T)\ \longrightarrow &\ W(\phi _j,T)\ \delta (icT\ +\ d)^{-3},
\cr}  \eqn \trans $$
where $\delta $ is a phase which is irrelevant for our purposes.
One says that the superpotential transforms as a modular
form of weight $-3$ and the matter field $\phi _j$ transforms
as a modular form of weight $n_j$.

Apart from its theoretical importance, duality symmetries
are important because the effective low-energy Lagrangian
has to respect this symmetry at the tree level and in loops.
This invariance has been checked explicitly  at the tree
level \LMN \ in orbifold compactifications and also in some
one-loop computations \DKLB ,\ANT  . More importantly, there are
reasons to believe that the symmetry is also $respected$
$by$ $non$ $perturbative$ $effects$, i.e. it may be broken
spontaneously but $not$ explicitly  \FLST ,\FILQ . This has important
consequences since e.g., if some non-perturbative superpotential
is generated, it will have to be a modular form of weight
$-3$. Since the type of modular forms available is very
limited, one can use this information to guess the form
of the possible non-perturbative effects \FILQ \ (see below).
One can also impose     one-loop invariance under modular
transformations, i.e. cancellation of $duality$ $anomalies$,
in order to constrain  the possible massless multiplets
present in a given model \IL   .
Thus, for example, it  can be seen that one cannot possibly
build a   $Z_3$  or $Z_4$ model in which the only massless
fields with non-vanishing quantum numbers under the
standard model are those of the minimal SUSY model \IL   .
Such a model would necessarily have duality anomalies.
Owing to these facts, duality symmetries are important
          ingredients to trace the stringy origin
of a given low-energy effective field theory. We will
now show the importance of these symmetries in several
different phenomenological contexts.

\bigskip
\leftline{\bf 3. String unification of coupling constants}

\bigskip

As we mentioned above, gauge coupling constants are
unified in string theory even in the absence of
a unification group such as $SU(5)$. However, the
Planck scale boundary conditions depend on the values of
the $k_i$'s. If we have a string with some sort of grand
unification structure, one expects to have $k_3=k_2=3/5k_1$,
leading to the standard GUTs boundary conditions. Nevertheless
those boundary conditions are also possible in
direct $SU(3)\times SU(2)\times U(1)$ strings ($k_3=k_2=1$ is very
common in specific models whereas getting  $k_1=5/3$ is
typically non-trivial). The actual success of the standard
GUTs boundary conditions makes it  advisable  to  construct
   strings with that property.

Unlike GUTs, in which one $computes$ the unification mass
in terms of the crossing point of two running coupling
constants, in strings $we$ $do$ $know$ $the$ $string$
$unification$ $scale$ since it is related to the Planck mass
in a known way. In the ${\overline {MS}}$ scheme one finds \KAPLU
, \ELLISETAL \
              $M_{string}=0.7\times g_{string}\times 10^{18}$
GeV.  Below this scale the couplings are renormalized in the
usual way \GQW   .
The one-loop running gauge coupling constant of a          gauge
group $G_a$ is of the following form:
$${1\over g_a^2(\mu)}={k_a\over  g_{\rm string}^2}+{b_a\over 16\pi^2}
\log{M_{string }^2  \over\mu^2}+\Delta_a.\eqn\rgea
$$
Here $b_a=-3C(G_a)+\sum_{j  }       T(R_j)$ is the $N=1$ $\beta $-
function coefficient and $\Delta _a$ represent possible
threshold effects at the unification scale. Now, since we do
know the unification scale, by running coupling constants
down to low energies one can compute not only $\sin ^2\theta _W(M_W)$
but some other gauge coupling, e.g. $\alpha _3(M_W)$. If one
assumes the particle content of the minimal SUSY-SM and
the big desert hypothesis, one gets (neglecting threshold effects)
\ILR ,\ANTON   :
$$
\sin ^2\theta _W(M_W)\ =\ 0.218\ \ \ \ ;\ \ \  \alpha _3(M_W)\ =\
0.20\ .  \eqn \resu $$
Both numbers are several standard deviations away from the
measured values (${sin^2\theta _W}_{exp}=0.233\pm 0.0008$,
${\alpha _3}_{exp}=0.115\pm 0.007$). There are three main possibilities
         to explain this disagreement (apart from forgetting about
strings, which is obviously foolish): i) There is an intermediate
scale $     M_X  \sim  10^{16}$ GeV at which          a GUT
symmetry like $SU(5)$ or $SO(10)$ is at work. This has
the very same problems of the standard SUSY-GUTs that we
discussed above. It also requires the use of higher
      Kac-Moody levels ($k_5\geq 2$), which is both
technically complicated and phenomenologically
problematic \HIGH ,\HIGHK . ii) There are additional massless chiral
particles on top of the ones of the minimal SUSY-model.
This was considered in ref. \PLANCK \  and more recently, in
the context of strings, in \ANTON   .
iii) We stick to the minimal low-energy content of the
standard model, but the string threshold effects $\Delta _a$
are important and correct for the disagreement with
experiment \ILR . Notice that the string threshold gives
potentially large corrections since it involves
an infinite tower of massive states.

The second possibility is perfectly possible, although one will
eventually have to explain how and why the required extra particles
(and no others) remained light. From the point of view of
pursuing minimal things first, the third possibility would
be more attractive. In the field theory case     Weinberg  provided
a long time ago a formula \WEIN ,\LLR to compute the threshold effects in
usual GUTs (in the $\overline {MS}$ scheme):
$$
\Delta ^a_{FT}\ =\ {1\over 2}T^a(V)\ +\ 4T^a(F)\log ({{M_X}\over {M_F}})
\ +\ {1\over 2}T^a(S)\log ({{M_X}\over {M_S}}) \  ,
\eqn \ftt $$
where $V,F,S$ refer to vector, fermion, and scalar heavy particles,
respectively, and $T^a$  are the quadratic Casimirs of those
particles. As we can see, in order to compute the
threshold effects we need to know two things: i) The complete
spectrum of heavy particles and ii) their quantum numbers
with respect to the unbroken gauge group. If we want to do
something similar in string theory we will thus need a
detailed knowledge of the supermassive spectrum (i.e. the
complete one-loop partition function). Of course, this will
be something               very model dependent in general.
Such a computation was worked out in the case of
orbifold 4-D strings in ref \KAPLU   . These corrections
are small, as is usually the case in field theory.
However, in certain situations (when there are massive modes
whose mass explicitly     depends on the compactification
radius $R^2$), the threshold string corrections may be
very large. This comes about because the gauge kinetic
function $f$ (whose real part equalls   $1/g^2$)
gets dependent on the moduli $T_i$ at one loop
(it is just $=k_aS$ at the tree level).
This fact was already realized a long time ago, using
scale-invariance arguments which lead  one to expect
(for large $R^2$) one-loop-corrected $f$-functions \IN \  :
$$
f^a\ =\ k_aS\ +\ \epsilon _a\ T    \eqn \ilf $$,
where $\epsilon _a$ are small group-dependent coefficients.
It is clear that for sufficiently large $R^2=ReT$ such
loop corrections can be very important. More recently
these corrections have been computed (for all $T$ values)
in the case of orbifold 4-D strings, yielding the result \DKLB
$$
f^a\ =\ k_aS\ +\ {{b_a'}\over {8\pi ^2}}\ \log (\eta ^2(T)) \ ,
\eqn \fst $$
where $\eta (T)$ is the Dedekind function and $b_a'$
is given by \    \LOUIS ,\DFKZ
$$
b_a'\ =\ 3C(G_a)\ -\ \sum _j T(R_j)(3+2n_j) \ , \eqn \bpr $$
where $C(G_a)$ is the quadratic Casimir in the adjoint (e.g. $=N$ in
$SU(N)$) and $n_j$ are the modular weights of the matter fields.
The sum  in \bpr \ runs over all $massless$ fields with $G_a$
quantum numbers.
The Dedekind function admits a large $T$ expansion
$\eta (T)\simeq e^{{-\pi T}\over {12}}(1-e^{-2\pi T}+...)$
and in such a limit we recover eq.  \ilf \ with
$\epsilon _a\ \simeq -b_a'/(48\pi )$. The one-loop
$T$-dependent piece in \fst \ may be understood as a threshold
correction due to the string massive modes with a $T$-dependent
mass. One thus finds for the string threshold
corrections
$$\Delta_a(T,\bar T)={1\over 16\pi^2} b_a'
       \log (T_R|\eta(T)|^4) \ ,\eqn
\threorbi
$$
where $T_R=T+\bar T=2R^2$. The extra piece, $\log \ T_R$, is in
fact related to the massless fields. This extra dependence on $ReT$
originates in one-loop graphs involving only massless fields and it
is there \DFKZ ,\LOUIS \ because of the Kahler invariance of the
$N=1$ supergravity Lagrangian and also because of the
tree-level
explicit $T$ dependence of the matter kinetic terms in
eq. \kahl \ . It is worth noticing that $\Delta _a$ is
explicitly  invariant \FILQ \ under the modular transformations
\mod   . Indeed, the Dedekind function is known to transform
like $|\eta (T)|^4    \rightarrow |icT+d|^2|\eta   (T)|^4$, and
this is cancelled by the transformation of $T_R$. Thus the
modular one-loop non-invariance of the massless sector is
cancelled by the contribution coming from the heavy
modes.

              Anyway, it is clear that a term like that in eqn.
\threorbi \ can give rise to substantial contributions
and represent the  $leading$ $threshold$ $corrections$
for sufficiently large $T$ \ILR   . Moreover these corrections
$only$ $require$ $knowledge$ $about$ $the$ $massless$
$sector$ of the theory (i.e. the $b_a'$'s).
This is in contrast with the field theory case of
eq.\ftt \ , in which full knowledge of the
massive sector of the theory is required. These corrections can thus
be computed                                         in terms of the
(known) quantum numbers of the massless particles and
their (unknown, but restricted \IL  ) modular weights $n_j$.
(The origin of this fact is that massless and massive sectors
of the theory are connected by duality and that the massive
contribution needs to have a certain form in order
to cancel the duality non-invariance of the
massless sector one-loop contribution.)
Now, taking different linear combinations of coupling
constants one arrives  at the results \ILR
$$
{\sin^2\theta _W}(\mu) \ =\
{{k_2}\over {k_1+k_2}} - {{k_1}\over {k_1+k_2}}
{{\alpha_e(\mu)}\over {4\pi}}\biggl(  A\ \log({{M_{\rm string}^2}
\over {\mu ^2}}) - \ A'\
\log(T_R|\eta (T)|^4)\biggr)  \ ,
\eqn \sint$$
where $A$ is given by
 $
A\ \equiv \ {{k_2}\over {k_1}}b_1-b_2
$
and $A'$ has the same expression after replacing
$b_i\rightarrow   b_i'$;
                                           $\alpha _e$ is
the fine structure constant evaluated at a low-energy scale
$\mu $ (e.g. $\mu =M_Z$). In an analogous way one can compute
the low-energy value of the strong interactions fine-structure
          constant $\alpha _s$
$${1\over {\alpha _s(\mu )}}\ =\ {{k_3}\over {(k_1+k_2)}}
\biggl({1\over {\alpha_e(\mu)}}\ -\ {{1}\over {4\pi }}\ B\
\log({{M_{\rm string}^2}\over {\mu ^2}})\    - \
{{1}\over {4\pi }}\ B' \
\log(T_R|\eta (T)|^4)\biggr)\ ,   \eqn \alfs $$
where
$
B\ \equiv \ b_1+b_2-{{(k_1+k_2)}\over {k_3}}b_3
$
and $B'$ has the same expression after replacing
$b_i\rightarrow b_i'$.
Taking the standard values $k_3=k_2=3/5k_1$ and setting
$A'=B'=0$, one recovers the results in eq. \resu   . On the
other hand one can see if the measured values can be accommodated
       by choosing apropriate values for $A',B'$ (i.e.
modular weights $n_j$) and $T_R=R^2$. One can eliminate the
$T$-dependence from eqs. \sint   and \alfs   . Defining
$\gamma =B'/A'$ one finds that          reasonable results can be
obtained for $sin^2\theta _W$ and $\alpha _s$ provided
        \ILR
$$
   \gamma \ =\ {5\over 3}{\alpha _e}\left( {{({1/      {\alpha _s}^0}
-{1/      \alpha ^{exp} _s(\mu )} )}\over {({sin^2\theta _W^0}-
sin^2\theta ^{exp}_W(\mu )}}\right) \ ,\eqn \bbet
$$
where $sin^2\theta _W^0$ and ${\alpha _s}^0$ are given by \resu   .
Allowing for experimental errors one finds the numerical constraint
$2.2\leq \gamma \equiv B'/A'\leq 4.0$. In addition, in order
for the threshold corrections to have the correct sign one
needs $A'>0,B'>0$. If all these conditions are met, there is always
a value of $R^2$ such that one gets good agreement with
experiment (of course, one has to  check that $R^2$ is not
so large that the threshold corrections are too big
and even dominate the tree-level  coupling constant!).
Equation \bbet \ translates  through eqn. \bpr   into
a constraint on the modular weights $n_j$ of physical
quark, lepton,  and Higgs superfields which can be used
to constrain specific 4-D string models (see ref. \IL    ).
The reader can check that a possible satisfactory solution
is obtained if , for instance,   $R^2\sim 16$ (in string units)  and one
has family-independent modular weights
$n_Q=n_D=-1$; $n_U=-2$; $n_L=n_E=-3$ and $n_H+n_{\bar H}=-5$.
Many other possibilities exist depending on the
available modular weights in each orbifold model.
This latter point is crucial and one can see that many  possibilities
can be ruled out            in this way \IL  .
For example, if only the overall modulus $T$ is considered,
                 the minimal scenario  discussed above
is not possible for any $Z_N$ orbifold (the situation changes
if the three different $T_j$ planes and/or larger Kac-Moody
levels are considered) \IL   .

The conclusion of the above analysis is that
string threshold effects can make   the predictions for
$\sin ^2\theta _W$ and $\alpha _s$ fit to
experimental results for moderately large values of $R^2$,
provided an orbifold with the minimal particle content
and apropriate modular weights exist (of course,
             extra  particles without standard model
quantum numbers are always allowed). The other obvious alternative
\PLANCK , \ANTON    , is having a particular set of extra massless matter
fields . But even in this case the threshold corrections could be
important and it could well be that both things were present
in the actual case chosen by nature \IL   . Notice equations
\sint   ,\alfs \ are valid for arbitrary low-energy particle
content, and it is likely that both a modification of the
massless particle content $and$ the contribution of large
$T$-dependent threshold corrections were required to fit
the data without intruducing new intermediate mass scales
in the theory \IL    .

\bigskip

\leftline{\bf 4. Duality-invariant supersymmetry breaking}

\bigskip

Another relevant problem in which duality seems to play an
important role is that of supersymmetry breaking in string
models. Specific 4-D string models very often come with
extra ``hidden" gauge interactions which couple to   the
observed particles only gravitationally. It was suggested  \GAUGINO ,
\DIN \ that if this hidden sector is strongly interacting,
gaugino condensates $<\lambda \lambda >$ could form
and give rise to spontaneous supersymmetry breaking.
An  intuitive way to describe this low-energy phenomenon
is through a non-perturbative (recall $ReS=1/g^2$) superpotential of the
form $W(S)\ =\ e^{{3S}/      2b_H }$, where $b_H$ is the
$\beta $-function coefficient of the hidden gauge group $H$.
This mechanism seems       attractive, since the exponential
factor could in principle generate the required small
supersymmetry breakdown. However, it was soon realized
that this scenario as discussed in ref.\GAUGINO ,\DIN \ had several
problems. 1) There is no stable (non-trivial) minimum
for the dilaton field $ReS$. 2) It is not clear what dynamics
could fix the size  of the compactified variety  $ReT$ since $W(S)$
does not depend on $T$. Furthermore, from the discussion in
previous sections, it is also clear that a potential
like $W(S)=e^{{3S}/      2b_H }$ cannot possibly be
an effective superpotential from superstring theory.
       Indeed, such a superpotential is invariant under
duality, whereas we learned that any superpotential
should transform as a weight $-3$ modular form\ !

It has been recently realized \FILQ ,\DUAGAU \ that a
gaugino condensation mechanism for supersymmetry breaking can
be naturally formulated
                       in a manner consistent with
duality invariance. Furthermore, this provides     a solution to
the second problem above, i.e.  the dynamical determination
of the compactification scale $ReT$. Let us consider for
simplicity the case of a condensing gauge group without
chiral matter fields. In this case one has $b_H'=-b_H$ (see eq.
\bpr    ) and $f_H=S -  {{b_H}\over {8\pi ^2}}log(\eta ^2(T))$.
Now, the relevant effective non-perturbative superpotential
can be estimated to be \FILQ
$$
 W(S,T)\ =\  e^{{3f_H (16\pi ^2)}/      2b_H }\ =\ {{e^{{3S(16\pi ^2)}
/      2b_H }}\over {\eta ^6(T) }} \ .
 \eqn \supot
$$
This superpotential has now two interesting properties. First,
it explicitly depends on the $T$ field. The existence of such a
$T$ dependence of the gaugino condensation superpotential
was already remarked in refs. \IN ,\YO \  but in that case the
large $T$ limit
            formula for $f$, eq.\ilf ,   was used and the corresponding
scalar potential was unstable \YO   . The second interesting property is
that now $W(S,T)$ transforms as a modular form of weight $-3$
(recall that $\eta ^2(T)$ has modular weight one), as required
by duality invariance. Owing to the presence of the $\eta (T)$
function the scalar potential is now well behaved. If one
only keeps the fields $S$ and $T$ and assumes a general
non-perturbative superpotential of the form
$W(S,T)=\Omega (S)/\eta ^6(T)$, the obtained scalar potential
is \FILQ
$$
V\ =\ {1\over {S_RT_R^3|\eta (T)|^{12}}}\{ |S_R\Omega _S-\Omega |^2
\ +\ 3|\Omega |^2({{T_R^2}\over {4\pi ^2}} |{\hat G}_2|^2\ -\ 1)\}
\ , \eqn \sem
$$
where $S_R=2ReS,T_R=2ReT$ and $\Omega _S=\partial
 \Omega  /\partial S$.
Here the function ${\hat G}_2$ is the weight-2   Eisenstein
function which admits the expansion ${\hat G}_2\simeq
{{-2\pi }\over {T_R}}\ +\ {{\pi ^2}\over 3}\ -
\ 8\pi ^2e^{{-2\pi T}}\ -....$. From the form of the potential
one can already draw some general conclusions. In the decompactification
limit $T_R\rightarrow \infty $ (and its dual $T_R\rightarrow 0$),
the potential diverges ($V\rightarrow \infty $) since   the product
$T_R^3|\eta (T)|^4$ vanishes exponentially at both points, and
${\hat G}_2\rightarrow \pi ^2/3$ as $T_R\rightarrow \infty $.
Thus, if gaugino condensation takes place,
the $theory$ $is$ $forced$ $to$ $be$ $compactified$. The origin
of the divergence for $T_R \rightarrow \infty $ has a clear
physical meaning \CFILQ   : for large radius $R$ an infinite number
of Kaluza-Klein excitations (with masses $\sim n^2/R^2,\ n\in
{\bf Z}$) become massless and contribute to the effective
coupling constants, which in turn diverge.
Because of duality, the same arguments apply for $T_R\rightarrow 0$
replacing Kaluza-Klein excitations by winding modes.
Another interesting property of the potential is its periodicity
in $ImT$ (direct consequence of $T\rightarrow T+i$ invariance).
The minimum of such a potencial is at $ImT=integer$, leading
to purely $real$ soft SUSY-breaking terms in the effective theory.
This might be interesting in trying to understand the
CP-violating properties \IBLU \ of this effective theory.

In the above scheme,           supersymmetry is
spontaneously broken                     with
$m_{3/2}=|\Omega|/(S_R^{1/2}T_R^{3/2}|\eta |^6)$.
The size of supersymmetry breaking is essentially determined
by the value of $|\Omega (S)|$. It is not yet clear what
dynamics (such as      multiple gaugino condensation or
other mechanism)     could lead to the required hierarchically small
SUSY-breaking  and it is not clear either what could
make the           cosmological constant vanish.
Concerning the problem of determining the vev of the
dilaton field $S$, the possibility of having a new
kind of duality symmetry for the $S$ field is intriguing \SDUAL   .
This new $S-duality$ will relate strong to weak coupling
($S\rightarrow 1/S$) and contain a new discrete
Peccei-Quinn-type symmetry ($S\rightarrow S+i$).
If such a symmetry (or some generalization) was present in the
effective low-energy theory one would expect background values
$S_R^{-1}\sim g^2\sim 1$, not far away from the self-dual
point. This is not very different   from the estimations
of  the values of the gauge coupling constants at the unification
scale (notice that $g_{GUT}$ is of order $\sim 0.7$ if one extrapolates
the measured low-energy  values). In fact this would provide
an explanation for why indeed the measured coupling constants
are so close to one and not to any other, e.g. much smaller,
value.
It is
anyway  not unreasonable to think that,
                                 independently of the
dynamics which eventually fix the value of $<S>$, the
$T$-duality of the effective action will force the theory
to be compactified.
              Notice also  that once the $T$ field gets a
vev, duality is generically spontaneously broken \FILQ   .

In the above discussion I have considered the simplest
possibility of a matterless condensing gauge group and
focused only on the generic fields $S,T$. The more general case
with matter fields \LT \ and several orbifold moduli $T_j$ can
also be considered arriving at similar qualitative
properties. A further complication arises  in some cases
if the dilaton field $ReS$ gets mixed at one loop
with the $T_j$ fields in the Kahler potential  \DFKZ   .
Except for some limiting cases (the $Z_3$ and $Z_7$ orbifolds)
the above simplified discussion gives the correct qualitative
answer.

More generally \CFILQ\ one can simply consider a generic non-perturbative
superpotential of the form $W=\Omega (S) H(T)/\eta (T)^6$ in
which $H$ is some modular invariant function (up to constant
phases). This type of non-perturbative superpotentials
could arise  in more complicated situations (not necessarily
linked to the gaugino condensation mechanism) involving
some unknown non-perturbative dynamics.
If $H$ depends only on $T$, one can see that either $|H|=constant$
or else there will necessarily be singularities for finite
values of the field $T$. It is not clear what kind of
physical situations could lead to these singularities, but
one can      find particular expressions for $H(T)$ (polynomials
of the known absolutely modular invariant function $j(T)$)
which give rise to a vanishing cosmological constant \CFILQ \ .
On the other hand, if there are matter fields coupling to the
condensing gauge group in the gaugino condensation
mechanism, $H$ may explicitly depend on those matter fields.

\bigskip

\leftline{\bf 5. Soft SUSY-breaking terms and duality invariance }

It is well known that when supersymmetry is broken in      a
``hidden''  sector of the theory, the physical world
feels  the breaking of SUSY through the existence of
explicit SUSY-breaking soft terms (gaugino masses $M_{\alpha }$,
scalar masses $m^2$,
trilinear scalar couplings $A_j$, etc. \MALLOR ). In general
the form of those soft terms is expected to depend
on the way in which SUSY-breaking takes place and on the
particular 4-D string considered. If the idea of low-energy
       supersymmetry is correct, the      soft terms
should    eventually be measured at accelerators. It would thus
      be very important        to find theoretical
constraints  on  these terms coming from generic string
properties like duality in order to check the theory.
Even if one were unable to obtain model-independent
predictions for those quantities, knowing the relation
between given classes of 4-D strings and the resulting
soft terms would perhaps allow us to rule out (or rule in\ !)
large classes of models.

It turns out that in some simple situations something can be said
          about the soft SUSY-breaking terms in a
relatively model-independent way, at least within the
class of orbifold-like 4-D strings \IL   .  Consider, to
start with, the case of gaugino masses and take again the
simplified model with just $S,T$  as moduli fields.
Let us assume that supersymmetry breaking takes place
in this singlet $S,T$ sector, as in the scenarios
discussed in the previous section.
The gaugino masses $M_{\alpha }$ are then given by
$$
M_{\alpha }\ =\ f_SG^{-1}_{S^{*}S}h_{S^{*}}\ +\
f_TG^{-1}_{T^{*}T}h_{T^{*}} \ ,
\eqn \gaug
$$
where $h_{S^{* }}=G_{S^{*}}e^{G/    2}$ is the $S$
auxiliary field and $h_{T^{*}}$ is the corresponding
$T$ auxiliary field.  Using eq. \fst , one can easily
obtain an expression for each gaugino mass of the
form
$$
M_{\alpha }\ =\ k_{\alpha }\ M_0(S,T) +\ b_{\alpha }'\ M'(S,T) \ ,
\eqn \mgaus
$$
where $M_0,M'$ are gauge-group-independent quantities
which depend on the details of supersymmetry breaking.
The gauge-group dependence comes only \nextline through
$k_{\alpha }$ and $b_{\alpha }'$. Consider now
the phenomenologically interesting case with gauge
group $SU(3)\times SU(2)\times U(1)$. One can easily
write down combinations of gaugino masses in which the
dependence on $M_0,M'$ drops out. A particularly interesting
one is the following
$$
{{(M_1\ +\ M_2\ -{{k_1+k_2}\over {k_3}}\ M_3)}\over {
({{k_2}\over {k_1}}\ M_1\ -\ M_2)}}\ =\ {{B'}\over {A'}}
\ \equiv \gamma  \ ,
\eqn \ggam
$$
where $A',B'$, and $\gamma $ were defined in section 3. (Notice
that here $\gamma $ is given by a ratio of combinations of
$b'$s; we are $not$ necessarily imposing  the constraint
in eq. \bbet  \ which is required in order to explain the
experimental results for $sin^2\theta _W$ and $\alpha _s$
in terms of string threshold corrections. We are $not$ assuming
a large $ReT$ value either.)
{}From eq.
\ggam \ one can obtain the general constraint amongst
gaugino masses  (assuming the standard $k_{\alpha }s$) \IL
$$
M_1\ ({3\over 5}\gamma -1)\ -\ M_2\ (\gamma +1)\
+\ {8\over 3}\ M_3\ =0  \  .
\eqn \gaugsem
$$
This equation applies at the unification scale and
is authomatically satisfied by the standard ``minimal''
GUT constraints  $M_3=M_2=3/5M_1$ but allows for more
general possibilities depending on the model dependent
parameter $\gamma $. Thus if one wants to understand
the measured values of $sin^2\theta _W$ and $\alpha _s$
in terms of string threshold corrections one must
have, as we discussed in section 3, $2.2\leq \gamma \leq 4.0$
( one has $\gamma =25/7$ if one further imposses joining
of coupling constants at the minimal model value $M_X\sim
10^{16}$GeV). In models in which $b_{\alpha }'=-b_{\alpha }$
(as in the case of $Z_2\times Z_2$ orbifolds) one has
$\gamma =B'/A'=B/A$ and hence one can compute $\gamma $
by simply knowing the low energy field content. Thus, for example,
in the models of ref. \PLANCK \ and \ANTON  \  in which one adds
particular extra fields
                   in order to obtain adequate coupling
unification, one can explicitely compute \IL \ the values of $\gamma $.

A first lesson to be learned from the above discussion is
that, in string models, there can be non-universal contributions
to soft terms and, in particular, to gaugino masses.
I believe that eq. \gaugsem\ provides us with a
useful  parametrization for  departures from gaugino mass
universality, since a   measurement of $\gamma $ gives us
useful l information about the underlying string models.
Of course, we have considered above a case with only one
relevant modulus  $T$ and neglected the effect of matter fields
in supersymmetry breaking. More general situations in which the
above arguments still hold are considered in ref \IL   .

One can also obtain some interesting results for the soft
SUSY-breaking scalar masses $m^2_j$. Indeed, using the
Kahler potential of        eq. \kahl \ and assuming some
generic non-perturbative superpotential, one can compute \CFILQ \ the
corresponding scalar potential $V_{\phi }$ in the presence of matter
fields $\phi _j$. Expanding $V_{\phi }$ to quadratic order in the
matter fields one gets
$$
V_{\phi }\ =\ V_0(S,T)\ +\sum _j \ V_j\ \phi _j\phi ^{*}_j\ +...\ ,
\eqn \vphy
$$
where $V_0$ has a generic form as in eq. \sem \  and the sum  runs
over the matter fields. To first order on those fields, all  the
non-universal dependence of their masses comes through the
modular weights $n_j$ in eq.\kahl \ . After rescaling the
kinetic terms, one finds for the soft masses  \IL
$$
m^2_{\phi _j}\ =\ m^2_{3/2}\ +\ V_0\ +\
n_j\   {m_0}^2             (S,T)  \ ,
\eqn \mass
$$
where $m_{3/2}$ is the gravitino mass, $V_0$ is given by eq. \sem \
(and is essentially the cosmological constant), and ${m_0}^2$
is some model dependent mass parameter. As we did in the case of
gaugino masses one can take linear combinations and ratios
to isolate the dependence on the modular weights. Of course,
the situation becomes more complex if one considers the dependence
on other $T_j$ moduli and if the matter fields themselves
get involved in the process of symmetry breaking. Anyway, the
moral of eqs. \mass \ and \gaugsem \ is clear:  in generic
string models $SUSY$-$breaking$ $soft$ $terms$ $are$ $not$
$in$ $general$ $universal$ but vary from particle to particle.
Since       a certain degree of universality is required
from phenomenology (e.g.  suppression of flavour-changing
neutral currents), this may also lead to the elimination of
large classes of string models.

\bigskip
\leftline{\bf 6. Final remarks}
\bigskip

Four-dimensional strings constitute the most promising candidates
for unified theories of all the interactions. Their low-energy
       limits are described by         effective field
theory Lagrangians, which are invariant under certain
symmetries including gauge and $N=1$ supergravity. Of
special interest are the duality-type symmetries, which
are intrinsically stringy in character and which seem
to be  present in all 4-D strings studied up to know.
They are important because they are suposed     to be
respected even by non-perturbative interactions
hence allowing us to shed some light on  the non-perturbative
dynamics which are supposed  to break supersymmetry and fix the
values of the vev's of the moduli and dilaton in the theory.
We have shown above how in a large class of models
(Abelian $(0,2)$ orbifolds) the duality symmetry plays
an important role in understanding some phenomenologically
interesting questions such as gauge-coupling unification,
supersymmetry breaking, and the nature of soft SUSY-breaking terms.
Some specific string models and/or particle content
assumptions may be ruled out by using these duality arguments.

 We think that
       it is       important to keep    advancing in
determining the form of the effective low-energy Lagrangian
from 4-D strings  and, in particular, the generalized duality
symmetries present in non-orbifold models.
With this knowledge one should,   in principle, be  able to
constrain  in a substantial way the expected form of the
effective Lagrangian. If the idea of low-energy supersymmetry
is correct, future accelerators will detect and measure in
detail the masses of the  sparticles and hence important
experimental information on the SUSY-breaking terms will be
available. In the context of 4-D strings, these soft
terms encode important information on the structure of
the underlying theory. It will then be  an important challenge
to learn to read this important information from these
experimental data and  to draw the corresponding conclusions
concerning the underlying string theory.

\bigskip
\leftline{\bf Acknowledgements}

I thank M. Cvetic, A. Font, D. L\"ust, F. Quevedo and G.G. Ross
for a most      enjoyable collaboration and J. Louis and C. Mu\~noz
for comments on the manuscript.

\endpage

\par \penalty-400 \vskip\chapterskip
   \spacecheck\referenceminspace \immediate\closeout\referencewrite
   \referenceopenfalse
   \line{\fourteenrm\hfil REFERENCES\hfil}\vskip\headskip
   \input referenc.texauxil
   
\endpage

\vfill\eject\bye